# What makes a good app description

He Jiang, Hongjing Ma, Zhilei Ren, Jingxuan Zhang, Xiaochen Li
School of Software
Dalian University of Technology
Dalian, China
{jianghe, zren}@dlut.edu.cn
{hongjing_ma, jingxuanzhang, li1989}@mail.dlut.edu.cn

## ABSTRACT
In the Google Play store, an introduction page is associated with every mobile application (app) for users to acquire its details, including screenshots, description, reviews, etc. However, it remains a challenge to identify what items influence users most when downloading an app.

To explore users' perspective, we conduct a survey to inquire about this question. The results of survey suggest that the participants pay most attention to the app description which gives users a quick overview of the app. Although there exist some guidelines about how to write a good app description to attract more downloads, it is hard to define a high quality app description. Meanwhile, there is no tool to evaluate the quality of app description. In this paper, we employ the method of crowdsourcing to extract the attributes that affect the app descriptions' quality. First, we download some app descriptions from Google Play, then invite some participants to rate their quality with the score from one (very poor) to five (very good). The participants are also requested to explain every score's reasons. By analyzing the reasons, we extract the attributes that the participants consider important during evaluating the quality of app descriptions. Finally, we train the supervised learning models on a sample of 100 app descriptions. In our experiments, the support vector machine model obtains up to 62% accuracy. In addition, we find that the permission, the number of paragraphs and the average number of words in one feature play key roles in defining a good app description.

## Categories and Subject Descriptors
D.2.7 [**Software Engineering**]: Distribution, Maintenance, and Enhancement; D.2.9 [**Software Engineering**]: Management

## 1. INTRODUCTION
With the development of Android operating system, the number of available applications on Google Play increases very quickly. Some hot applications have more than 50,000 downloads in several weeks since they are launched [1]. To investigate what attracts users to download an app, we conduct a survey to find out which items users will consider when they download or install an Android application. There are totally 64 participants to share their opinions with us. From their perspective, the app description plays an important role in the items that the participants consider.

To drive more downloads, developers need to provide high quality descriptions about applications, because users usually know an application via its screenshots and description. Actually, there are some guidelines for developers to create a good description for encouraging users to download their applications. For example, some keys to a good description suggest to highlight the stand-out features and to sing the praises [2]. However, some tips about descriptions are not very concrete, such as "be smart", "be clear", "be informative", "be concise" [3].

In the guideline about "how to write an app description and drive more downloads (with examples)" [4], the importance to write app description well is pointed out. To generate maximum downloads, the developers had better ensure the app description is clear, brief and appealing to target audience. Because it is the app description that provokes users' curiosity besides the icon and name that help the application stand out.

In addition, there are some specific tips to optimize the descriptions. Laurie Galazzo [5] holds that it is important to find the best mix between content and form of the description and proposes his best practices to present the description. He suggests to use short sentences (+/- 120 characters per line), small paragraphs (+/- 3-4 lines per paragraph) , bullet points or lists, and Unicode symbols according to the app content & audience.

Besides, Microsoft [6] also provides some tips for writing a good, attention-grabbing description, like using lists and short paragraph, using a length that is just right and not including links or info that belongs elsewhere.

However, how do the users think about? Which attributes do the users care when they browse an app description in the Google Play store?

In this paper, we investigate the quality of app descriptions from the perspective of users. We try to find some attributes which impact the quality of app descriptions by the method of crowdsourcing. We invite some participants in our academy to rate the quality of app descriptions which are downloaded from Google Play. The requirement for participants is to evaluate the quality of app descriptions with a score ranging from one to five and explain the reasons for every score. Then, we extract 7 attributes from the reasons manually, namely the number of words, the number of features, the average number of words in every feature, the number of paragraphs, and presence of permission information, links and notes or tips. With all the attribute values and the average score of every app description, we train supervised learning models to evaluate the quality of app descriptions on a sample of 100 app descriptions. In our experiments, the SVM classification model achieves an accuracy of 62%.

| Q1: Have you had any experience of developing an Android app? |  |  |
|---|---|---|
| ○ No  ○ Yes |  |  |
| **No developing experience** |  |  |
| Q2: Which of the following items have you previously considered when downloading or installing an Android app? |  |  |
| Q3: Which three items affected you the most? |  |  |
| ☑ Category | ☐ Description | ☐ Rating |
| ☐ App Screenshots | ☐ Number of Rating Persons | ☐ Number of Installations |
| ☐ Size | ☐ Version | ☐ Reviews |
| ☐ What is New | ☐ Required Authority | ☐ Required Android Version |
| **Having developing experience** |  |  |
| Q4: Which of the following items have you previously provided when releasing an Android app? |  |  |
| Q5: Which three items were the most important to provide? |  |  |
| Q6: In your opinion, which three items are most crucial for users when downloading or installing an Android app? |  |  |
| ☐ Category | ☐ Description | ☐ Rating |
| ☐ App Screenshots | ☐ Version | ☐ Required Android Version |
| ☐ What is New | ☐ Size | ☐ Required Authority |

**Figure 1. The Questionnaire**

The remainder of this paper is organized as follows. Section 2 presents the details of our survey while section 3 illustrates the process of extracting attributes by crowdsourcing. Section 4 analyzes how to measure the quality of app descriptions and section 5 points out threats to validity. Finally, the related work and the conclusion are discussed in Section 6 and Section 7, respectively.

## 2. SURVEY
### 2.1 Survey Design
We design a survey to find the items that users will take into account when downloading applications. We put forward our questions in the questionnaire (see Figure 1).

As many participants have the experience of developing an Android app, we cannot ensure they have the same perspective with those having no experience about Android developing. Therefore, we need firstly confirm whether the participants have developed some Android applications, then we request them to answer different questions.

For the participant who has no developing experience, he/she needs to answer two questions: 'Q2-*Which of the following items have you previously considered when downloading or installing an Android app?*' and 'Q3-*Which three items affected you the most?*'. All the items of Q2-3 are provided in the Google Play store. The participants can select many items for Q2 but three items at most for Q3. According to Q2, we want to know whether all the items are valuable. Q3 helps us to explore the important items for users.

However, if the participant has experience of developing, he/she is requested to answer the questions Q4-6. The items for Q4-5 are short of 'Number of Rating Persons', 'Number of Installations' and 'Reviews' compared to the items of Q2-3, because it is impossible to provide the three items for a new application. By the two questions of Q4-'*Which of the following items have you previously provided when releasing an Android app?*' and Q5-'*Which three items were the most important to provide?*', we can check whether developers provide all the items that users consider. We also set Q6-'*In your opinion, which three items are the most crucial for users when downloading or installing an Android app?*', specially, the items for Q6 are the same as Q2-3. Similarly, participants can select many items for Q4 while at most three items for Q5-6.

We expect to find the important items for users by Q3 and Q6. We invite the participants in our academy to finish the survey by releasing the survey link[1].

### 2.2 Survey Results
There are totally 64 participants who response the survey, out of which 49 have no experience about Android developing, while others have. We analyze the results of Q3 and Q6, the percentage of every item in questionnaire is presented in Figure 2.

For 49 participants with no developing experience, 61% of them will consider the item of description. Besides, among those who have the experience of developing, there are 49% of the participants will consider the descriptions of Android applications.

---
[1] http://kwiksurveys.com/s.asp?sid=uqgwvhsjj2neqif361409

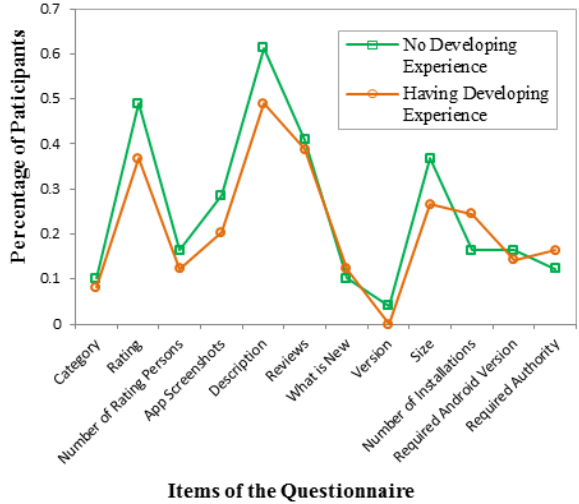

**Figure 2. Percentage of every item in questionnaire**

Moreover, the results suggest that the top three items that the participants care most are the description, the rating and the review of an application. However, for a newly released application, which has no information about rating and reviews, its description plays an important role in presenting its functions.

As a result, developers need to provide the good app description to drive more downloads. However, there are not any criteria for a high quality app description. We attempt to find some attributes to evaluate its quality and then provide some practical advices for the app developers.

## 3. EXTRACTING ATTRIBUTES WITH CROWDSOURCING

To find the attributes that users consider important when they scan the app descriptions in the Google Play store, we employ the method of crowdsourcing. The flowchart of rating by crowdsourcing is presented in Figure 3. We first download 50 app descriptions selected randomly from Google Play, namely app descriptions[C]. The second step is to invite users to evaluate the quality of app descriptions by means of rating and explaining the reasons of every score. Then we analyze the reasons manually to extract the common attributes that most users state. Next, based on the attribute list, we compute the values of all the attributes to get the attribute vector.

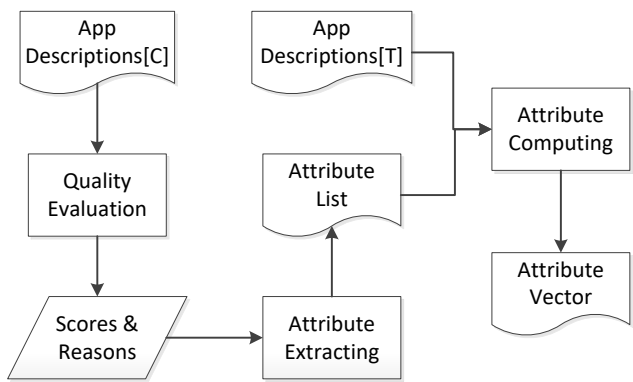

**Figure 3. Flowchart of Rating by Crowdsourcing**

### 3.1 Description Preparing

To prepare the app descriptions, we first confirm five app categories to download, namely Music & Audio, Photography, News & Magazines, Travel & Local, and Weather.

For every category, we download 10 app descriptions. Every description is saved as a text file with its category and name.

### 3.2 Crowdsourcing and User Rating

After preparing the description files, we aim to collect the users' feedback for app descriptions' quality. For this goal, we consider that the method of crowdsourcing is appropriate since crowdsourcing attracts more and more attention in the software engineering research [9, 10].

The definition of crowdsourcing was proposed by Jeff Howe and Mark Robinson [11] in the June 2006. It is often undertaken by sole individuals. Besides, the crucial prerequisite is the use of the open call format and the large network of potential laborers.

Kathryn [12] explored the use of crowdsourcing to support empirical studies in software engineering as it is a major challenge to evaluate a technique or tool on a large scale. Given this, we decide to rate the quality of app descriptions by crowdsourcing.

To evaluate the 50 app descriptions, we set the task to be that every description should be rated five times. 10 participants who major in software engineering of Dalian University of Technology accept this task. We believe that all the participants are adequate for this task. Each of them needs to evaluate an average of 25 app descriptions. Then every participant receives 25 text files about app descriptions. We also provide a rating text file for each app description, which is unitized with the keywords of category, name, rating, and reasons. Moreover, a task introduction file is attached in the email. In this file, we illustrate the purpose and deadline of the task, together with the request of score which ranges from one (very poor) to five (very good), and can be decimals, the higher the score, the better the quality.

### 3.3 Results Analysis

#### 3.3.1 Reasons Analysis

There are totally 250 scores due to five scores of every description, and 698 reasons provided by all the participants. Figure 4 presents the number of reasons and average score of every participant. In terms of quantity, the number of reasons participants provide varies from 42 to 91. Certainly, there are lots of personal factors in the results. However, it reflects the diversity of results. As the average score of every participant, some participants rate higher while some lower, but there is not obvious gap.

Similarly, Figure 5 demonstrates the number of reasons and average score of every description. On the whole, the average scores of the descriptions are related with the number of reasons while the higher the score, the more the number of reasons.

#### 3.3.2 Scores Analysis

Figure 6 plots the distribution of all the scores rated by the participants. As the figure shows, the x-axis means # of app descriptions while the y-axis represents the score of every description. As we said before, every description has 5 scores. From the distribution of all scores, most of scores range from 3 to 5, only 8 (3.2%) scores are less 1.5. There are a few app descriptions whose scores are more diverse, such as # 5, whose highest score is close to 5 while lowest score is only 1. The reason might be that this app description contains a controversial attribute. However, for the majority of app descriptions, there does not exist large gap among the 5 scores. Especially, #31, #39, #40 and #50 of the app description have very similar scores.

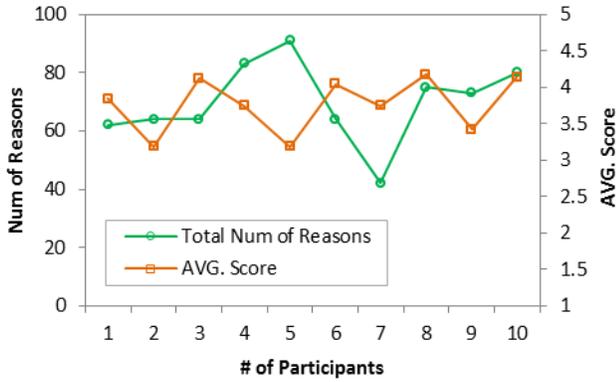

Figure 4. Rating Results of Every Participant

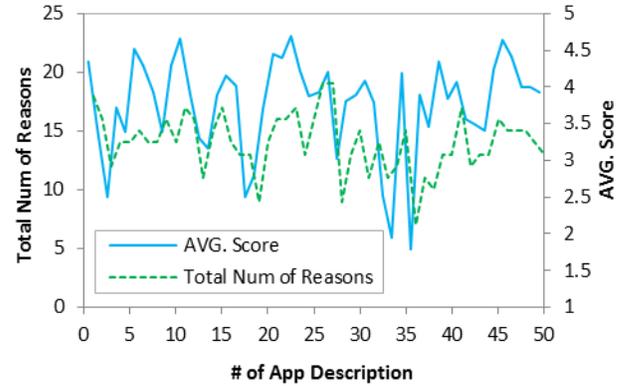

Figure 5. Rating Results of Every App Description

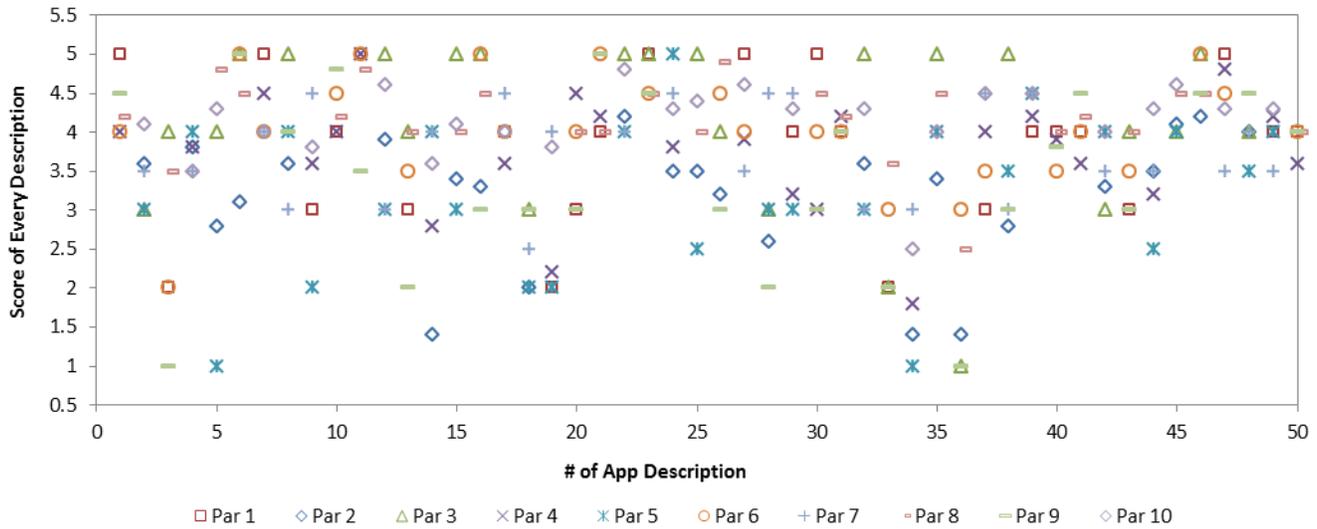

Figure 6. Distribution of Ratings by Participant

In addition, the highest average score in all the app description is #23, which gets two 5-points, and three 4.5-points with the 4.7 average score. Meanwhile, the #36 app description achieves the lowest average score of 1.78. Three of the five points are less 1.5 and one point is 2.5 and the other is 3. Actually, its description contains only one sentence.

### 3.4 Attribute Extracting

We expect to extract some attributes which have an influence on app descriptions' quality in the perspective of participants. For this aim, we manually analyze the 698 reasons by merging the similar meaning of sentences. In addition, we eliminate those reasons which are ambiguous or meaningless, for example, sentence likes this "*I don't understand this app is used for, it seems to be not just a music player.*"

The reasons are very abundant and cover a plenty variety of aspects. Most of reasons stress the format and content of the description. However, there are some reasons related to the category of the app. All the attributes are listed below:

**The structure of app description**. "*The structure is clear.*" "*It is organized in order.*" "*The structure is not good, it is hard for users to find what they what to see.*"

**The functions of application**. "*The description introduces its main functions.*" "*It contains some typical features.*" "*It lists its main function.*" "*The function is too single and not innovative.*" "*It is good for RSS reader to be small and quick, but not enough powerful for functions.*" "*The function of smart magazine is so attractive, but can be depicted more highlighted.*

**The length of description**. "*It is too long.*" "*The length is moderate as a whole.*" "*The content is tediously long, many words make no sense.*" "*Too many words, no patience to read on.*" "*It is so lengthy that it generates a bad influence on the understanding of readers.*" "*It is too short so that I have no idea how to use.*"

**Information about permission or requirement**. "*The information about permission is quite important while some users will worry about their privacy related to some higher permission.*" "*Requirement is Android 4.4+. Though this information is not necessary as version requirement is generally Android 2.3, it is better to declare if the version requirement is not traditional.*"

**Contact links**. *"Are the final links useful? The description contains usually app's function and feature." "It is so brief and provides only links, why not extract the content of links as the summary?" "I feel most of links in description are useless, especially the link for registering."*

**Note or tips**. *"The tips are so nice." "It contains tips and kindly note." "I like the final note." "The note about device improves the score."*

**User feedback**. *"What the experts said is too much. It is better to select some representative point. Going too far is as bad as not going far enough." "There are so many quotation marks, which cite what others say. How confusing!"*

**FAQ or how to use**. *"The content about how to use should not be put into the description, it can be added into the application." "I am not interested in 'faq' and 'visit us', I will not click it."*

**Functions related to app's category.** *"It will recommend personalized music based on individual music taste." " News can be read offline." "It supports credit card payment when reserving hotels."*

## 4. MEASURING APP DESCRIPTION QUALITY

In section 3.4, we state some attributes found from participants' reasons. We expect that these attributes are factors that influence the quality of app description. Therefore, we conduct some experiments of supervised learning models to explore our conjecture.

### 4.1 Description Preparing

As we extract attributes from the 50 app descriptions, we download another 100 app descriptions again from Google Play to avoid the attributes overfitting. We select randomly 20 app descriptions in every category which is the same as before. Similarly, every app description needs to be rated five times. Based on this premise, we divide randomly 25 different app descriptions into one group which is assigned to one participant.

### 4.2 Input Attribute Value

To build supervised learning models, the feature vector which represents every description should be composed of every attribute value. To obtain these values of attributes, some simple rules for calculating are set. For every attribute, the relative value is either binary or numerical.

Based on those attributes that section 3.4 mentions, the attributes we finally use are listed below. We discard some attributes which are hard to calculate, such as some specific functions related to the category of applications.

**The number of paragraphs**. Generally, participants do not like the descriptions in which there are some paragraphs that contain too many words. Besides, the number of paragraphs reflects the structure to some degree. In the website of Google Play, the paragraphs are different from newlines by different tags, while in the text files, we replace the tag of paragraph with double newline characters. As a result, we recognize the different paragraphs by splitting the description by the two line breaks.

**The number of features**. Whether the description contains features and whether these features are listed in an organized way play an important role in the participants' first impression.

In general, the app description introduces its functions or features with the keyword 'features' followed by a newline. For this reason, we first detect whether the description accords with the principle. If the description does not contain the keyword followed by a newline character, this attribute will be 0. Otherwise, we calculate the number of features by recognizing the itemization in the content below keyword 'features'. We try to examine whether several continuous lines start with same character or digit. However, there are some descriptions that highlight their features by capitals which are hard to distinguish. As we do not find reasonable rules to guarantee the accuracy of this condition, the type of capitals is ignored.

**The average number of words in every feature**. Compared to those long features, the features whose length is moderate will inspire participants a higher interest in reading. Usually, the length of every feature in one description are similar, we take the average number of words in every feature as an attribute. As the computing method, we combine the approach of calculating number of features with the process of counting words. Firstly, recognize the content of features, then get the total number of words of this content, finally, gain the average words of every feature by using total number of words to divide by number of features.

**The number of words**. To measure the length of description, we count the number of words. At the beginning, we replace all the characters of newlines with blank spaces. Then we make use of regular expression starting with digits and letters and ending with blank space to recognize all the words. We should note that some links will be regarded as one word.

**Permission**. We identify this attribute with regular expression which includes the keyword 'permission' and ends with newline. We only examine whether there is some permission information in the description while ignoring the content and the quantity of permission.

**Links**. It is easy to recognize the links with regular expression. All the links have the same prefix of 'http://' or 'https://'. As not all the descriptions have this attribute, we record this information by binary value, namely, if the description contains some links, the value of this attribute is 1, otherwise, the value is 0.

**Tips or Note**. The description that contains some tips will acquire a higher score according to the reasons from participants. We identify this attribute with regular expression starting with the keyword 'tips' or 'note' or 'notice' and ending with colon, moreover, the keyword 'please' is taken as an option in the regular expression.

### 4.3 Output Quality Level

We take the quality level of app descriptions as the class labels to train the supervised learning models. We confirm the quality level based on the average score of every description's 5 score rated by the participants.

To collect the scores of the 100 app descriptions, we invite 20 participants in our academy to finish the task. The participants are demanded to rate the app descriptions on a five-point scale [7] ranging from one (very poor) to five (very good), and decimals permitted. All the scores of each participant are recorded in one rating text file. Moreover, a brief introduction about the task is declared in the attention files.

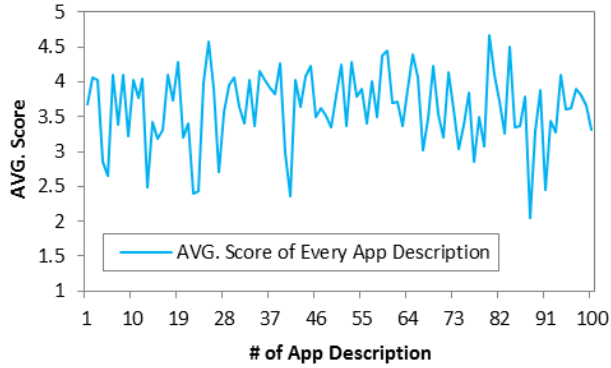

**Figure 7. The Average Score of Every App Description**

The average score of every app description is presented in Figure 7. Every point in the figure means the average score of the app description. From this curve, the scores are relatively centralized. Considering this imbalance that may cause the classification models to underperform, we refer to the previous work [13] to balance the data. We sort all the scores in descending order, then select two score points to divide all the scores into three quality levels – good, neutral, bad as equally as possible.

Table 1 shows the score scale and sum of app descriptions in every quality level. The description whose average score is higher than 3.9 belongs to good. However, the score lower than 3.4 is regarded as poor.

The quality level of every description is used for the class label in the next classification models.

**Table 1. Score Scale and Sum of Different Quality Level**

| Quality Level | Scale of Score | Sum of App Description |
|---|---|---|
| Good | (3.9, 4.66) | 33 |
| Neutral | (3.4, 3.9] | 33 |
| Bad | [2.04, 3.4] | 34 |

### 4.4 Evaluation

The 100 app descriptions rated only with scores are taken as the sample of our experiments to train the supervised learning models, which are used for evaluating the quality of app descriptions. Referring to the previous related work [13], we use these models: Support Vector Machines (SVMs), decision tree and random forest.

For SVMs models, we use the tool LIBSVM developed by Chih-Chung Chang and Chih-Jen Lin [14]. This tool realizes both support vector classification and regression. Meanwhile, it supports multi-class classification. In addition, we run the experiments of decision tree and random forest with Weka [15].

In a series of SVM classification experiments, the parameter of cross-validation is set 10. Similarly, the option of decision tree and random forest is 10-fold cross-validation.

Table 2 presents the accuracy of SVM classification models with different kernel, while Table 3 shows the mean squared error (MSE) of SVM regression models.

In Table 2, the numbers in right column indicate the percentage of prediction accuracy of relative classification models. It is obvious that the best model is SVM classification with radial kernel while

**Table 2. Prediction Accuracy of Classification Models**

| Model | Accuracy (%) |
|---|---|
| SVMC: radial kernel | **62** |
| SVMC: sigmoid kernel | 45 |
| SVMC: linear kernel | 47 |
| Decision tree | 40 |
| Random forest | 43 |

*SVMC = Support vector machines classification

**Table 3. Results of SVMR Model**

| Model | MSE |
|---|---|
| SVMR: radial kernel | **0.1754** |
| SVMR: sigmoid kernel | 0.2600 |
| SVMR: linear kernel | 0.2260 |

*SVMR = Support vector machines regression

the performance of other models is similar. On a whole, SVM classification models obtain better results than the other two models.

Moreover, Table 3 presents the results of regression model. Based on the premise that the smaller the MSE, the better the results, the SVM regression model with radial kernel performs best. This conclusion accords with the results of SVM classification models. From the MSE of SVM regression models, we analyze that there are some scores in some quality level and very close to the adjacent quality level to be classified by mistake.

### 4.5 Importance of Attributes

To check the effectiveness of the attributes, we investigate the importance of every attribute with the C4.5 decision tree with the Weka implementation J48 [16].

Figure 8 presents the decision tree for the app descriptions. The most important the attribute is, the higher situation it will place in the tree. The inner nodes in the tree are decision nodes which are aligned with vertical lines. Leaf nodes signify this path is over with the quality level classified into. After the quality level, there are two numbers in the parentheses. The first means the sum of app descriptions classified by this path while the second number explains how many app descriptions are assigned to the wrong class. The second will be omitted if the value is zero.

For example, if one app description has these attributes: permission information and the number of paragraphs is less than or equal to 19, the average words in one feature is less than or equal to 11.1429, no tips and no links, it will be labeled as bad quality.

From Figure 8, the three most important attributes are the permission, the number of paragraphs (para_num) and the average number of words in one feature (feature_avg_words). However, the app description which has no permission is classified as good. This result is contrary to what we expect. We speculate that the permission information may have an influence on other attributes

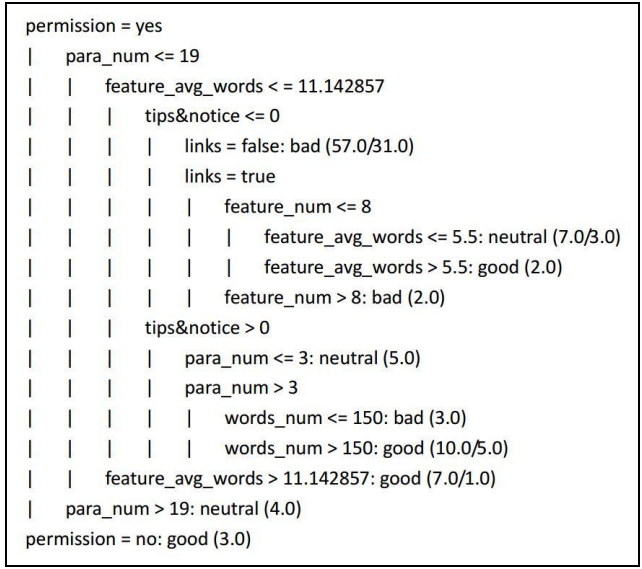

```
permission = yes
|   para_num <= 19
|   |   feature_avg_words <= 11.142857
|   |   |   tips¬ice <= 0
|   |   |   |   links = false: bad (57.0/31.0)
|   |   |   |   links = true
|   |   |   |   |   feature_num <= 8
|   |   |   |   |   |   feature_avg_words <= 5.5: neutral (7.0/3.0)
|   |   |   |   |   |   feature_avg_words > 5.5: good (2.0)
|   |   |   |   |   feature_num > 8: bad (2.0)
|   |   |   tips¬ice > 0
|   |   |   |   para_num <= 3: neutral (5.0)
|   |   |   |   para_num > 3
|   |   |   |   |   words_num <= 150: bad (3.0)
|   |   |   |   |   words_num > 150: good (10.0/5.0)
|   |   feature_avg_words > 11.142857: good (7.0/1.0)
|   para_num > 19: neutral (4.0)
permission = no: good (3.0)
```

**Figure 8. Decision tree**

such as the number of paragraphs (para_num) and the number of words (words_num). Meanwhile, the data in our experiments are limited.

However, there are some inspiring results, namely the app description that has the information about permission and tips or notice and the para_num is more than 3 will be classified as good if the words_num is more than 150, otherwise, the quality level is bad. We hope that this can provide some suggestions for the app developers.

## 5. THREATS TO VALIDITY

We identify some threats to validity from three aspects which include the selection of participants, the sample of app descriptions and the calculation of attributes.

Firstly, all the participants of survey and rating are from our academy. As we major in Software Engineering, what we require about the app descriptions may be different from the others with no knowledge related to computers. Furthermore, it is possible that someone approval an attribute which the others oppose. To reduce the influence of personal factors, we randomly choose 5 persons to rate one app description and take the average score to represent the quality of an app description.

Secondly, the applications in our experiments are from 5 categories. We extract attributes from the reasons of all app descriptions, but we do not discuss whether the attributes differ from the category of applications. What's more, the sample in our experiments contains totally 100 app descriptions, which are so limited that we use the cross-validation to achieve the reliable and stable models.

Thirdly, when analyzing the reasons and extracting attributes, we find some controversial attributes, such as praises from experts and users which some participants approve while some not. Besides, the opinion that the participants hold about links and FAQ (Frequently Asked Questions) is inconsistent. For this situation, we just use the links attribute which is easy to be identified by regular expression. As other attributes we use in our experiments, we calculate their values according to some regulars. To cover as more as app descriptions, we make the regulars as applicably as possible. However, there are still some flexible app descriptions which are omitted. For this issue, we will collect more app descriptions to find more reasonable regulars for avoiding the values of some attributes being disturbed in the future.

In addition, we wonder whether these attributes are suitable to the app descriptions in other platforms, such as apps for iPhone and windows phone.

## 6. RELATED WORK
### 6.1 Measuring bug report quality

One of the most related works is the research investigated by Thomas Zimmermann et al. [7, 13] about what makes a good bug report. To find out the factors that affect the quality of bug reports, they first revealed a mismatch between what developer need and what users supply, then they extracted those important features the majority of developers approve by analyzing their feedback. In addition, they developed a prototype tool called CUEZILLA to evaluate the quality of bug reports and provide some suggestions about how to improve the quality of bug reports.

### 6.2 Constructing feature models from product descriptions

There are already some studies about the description of product. Jean-Marc Davril et al. [17] constructed Feature Models (FMs) from product descriptions by an automated approach. They held that though individual product description contains only a partial view of features in the domain, a large amount of descriptions can cover comprehensive features.

Horatiu Dumitru et al. [8] mined product descriptions from publicly available online specifications and utilized text mining and clustering algorithm to discover domain-specific features, and generated a probabilistic feature model for on-demand feature recommendations. They validated their approach in 20 different product categories with thousands of product descriptions.

### 6.3 App Security

In addition, Alessandra Gorla et al. [18] tried to check app behavior against app descriptions by their prototype CHABADA. They identified outliers whose API usage is different from other apps in the same cluster by clustering Android apps descriptions' topics.

In contrast, Jialiu Lin et al. [19] demonstrated a new way to evaluate mobile app's privacy. They explored user's mental models of mobile privacy by crowdsourcing and achieved some interesting findings.

## 7. CONCLUSION

Android applications are attracting more and more downloads in recent years. It is convenient for users to know an application in Google Play store, which offers an introduction page for every application. Among all the detailed items about the applications, we wonder which items users consider more when they download an app. Therefore, we conduct a survey to explore the important items in the perspective of users. From the survey results, we find that more than half of the participants will pay attention on app descriptions. Undoubtedly, a high quality app description can drive more downloads for an app. Though there are some guidelines about how to improve the quality of app description, most of them are not practicable. We try to find out some attributes that affect the quality of app descriptions. Firstly, we download 50 app descriptions from Google Play. Then we invite some participants to rate the quality of app description with the method of crowdsourcing. When they rate the descriptions, they are requested to explain the reasons of every score. By analyzing these reasons, we extract some attributes to measure the quality of app descriptions.

We take the attribute values as the input while the quality level as the class label to train supervised learning models. We calculate every attribute value based on some regulars that we make. We partition the quality level based on the average score of app descriptions. We use a sample of 100 app descriptions to validate the performance of evaluating the quality by these attributes.

In our experiments, the SVM classification model with radial kernel achieves the best results of an accuracy of 62%. Additionally, in order to inspect the importance of every attribute, we conduct an experiment based on the C4.5 decision trees. According to the results, the permission, the number of paragraphs and the average number of words in one feature are the most three important attributes for app descriptions. We appreciate some results of the decision tree, hoping that it can provide some suggestions for the app developers.

## 8. ACKNOWLEDGMENTS

Sincerely thanks to all the participants who major in software engineering in Dalian University of Technology to response the survey and evaluate the quality of app descriptions. This work was supported in part by the New Century Excellent Talents in University under Grant NCET-13-0073, and in part by the National Natural Science Foundation of China under Grant 61370144.

## 9. REFERENCES

[1] Downloads per app since launch date. http://www.appbrain.com/stats/android-app-downloads.

[2] App descriptions for the app store. http://www.ibabbleon.com/app-store-app-description-writing.html.

[3] 5 simple steps to the perfect app store description. http://www.pocketgamer.biz/feature/52922/5-simple-steps-to-the-perfect-app-store-description/.

[4] How to write an app description and drive more downloads (with examples). http://www.trademob.com/how-to-write-an-app-description-and-drive-more-downloads-with-examples/.

[5] 5 tips to optimize your app description. http://blog.apptweak.com/2014/05/5-tips-to-optimize-your-app-description.

[6] Your app's description – Windows app development. http://msdn.microsoft.com/en-us/library/windows/apps/hh694076.aspx.

[7] Bettenburg, N., Just, S., Schröter, A., Weiss, C., Premraj, R., & Zimmermann, T. (2008, November). What makes a good bug report?. In Proceedings of the 16th ACM SIGSOFT International Symposium on Foundations of software engineering (pp. 308-318). ACM. DOI= http://doi.acm.org/10.1145/1453101.1453146.

[8] Dumitru, H., Gibiec, M., Hariri, N., Cleland-Huang, J., Mobasher, B., Castro-Herrera, C., & Mirakhorli, M. (2011, May). On-demand feature recommendations derived from mining public product descriptions. In Software Engineering (ICSE), 2011 33rd International Conference on (pp. 181-190). IEEE. DOI= http://doi.acm.org/10.1145/1985793.1985819.

[9] Stol, K. J., & Fitzgerald, B. (2014, June). Researching crowdsourcing software development: perspectives and concerns. In Proceedings of the 1st International Workshop on CrowdSourcing in Software Engineering (pp. 7-10). ACM. DOI= http://doi.acm.org/ 10.1145/2593728.2593731.

[10] Musson, R., Richards, J., Fisher, D., Bird, C., Bussone, B., & Ganguly, S. (2013). Leveraging the crowd: how 48,000 users helped improve Lync performance. Software, IEEE, 30(4), 38-45. DOI= http://doi.acm.org/ 10.1109/MS.2013.67.

[11] Brabham, D. C. (2008). Crowdsourcing as a model for problem solving an introduction and cases. Convergence: the international journal of research into new media technologies, 14(1), 75-90. DOI= http://doi.acm.org/10.1177/1354856507084420.

[12] Stolee, K. T., & Elbaum, S. (2010, September). Exploring the use of crowdsourcing to support empirical studies in software engineering. InProceedings of the 2010 ACM-IEEE International Symposium on Empirical Software Engineering and Measurement (p. 35). ACM. DOI= http://doi.acm.org/10.1145/1852786.1852832.

[13] Zimmermann, T., Premraj, R., Bettenburg, N., Just, S., Schroter, A., & Weiss, C. (2010). What makes a good bug report?. Software Engineering, IEEE Transactions on, 36(5), 618-643. DOI= http://doi.acm.org/10.1109/TSE.2010.63.

[14] Chang, C. C., & Lin, C. J. (2011). LIBSVM: a library for support vector machines. ACM Transactions on Intelligent Systems and Technology (TIST),2(3), 27. DOI= http://doi.acm.org/ 10.1145/1961189.1961199.

[15] Hall, M., Frank, E., Holmes, G., Pfahringer, B., Reutemann, P., & Witten, I. H. (2009). The WEKA data mining software: an update. ACM SIGKDD explorations newsletter, 11(1), 10-18. DOI= http://doi.acm.org/10.1145/1656274.1656278.

[16] Witten, I. H., & Frank, E. (2005). Data Mining: Practical machine learning tools and techniques. Morgan Kaufmann.

[17] Davril, J. M., Delfosse, E., Hariri, N., Acher, M., Cleland-Huang, J., & Heymans, P. (2013, August). Feature model extraction from large collections of informal product descriptions. In Proceedings of the 2013 9th Joint Meeting on Foundations of Software Engineering (pp. 290-300). ACM. DOI= http://doi.acm.org/10.1145/2491411.2491455.

[18] Gorla, A., Tavecchia, I., Gross, F., & Zeller, A. (2014, May). Checking app behavior against app descriptions. In ICSE (pp. 1025-1035). DOI= http://doi.acm.org/10.1145/2568225.2568276.

[19] Lin, J., Amini, S., Hong, J. I., Sadeh, N., Lindqvist, J., & Zhang, J. (2012, September). Expectation and purpose: understanding users' mental models of mobile app privacy through crowdsourcing. In Proceedings of the 2012 ACM Conference on Ubiquitous Computing (pp. 501-510). ACM. DOI= http://doi.acm.org/10.1145/2370216.2370290.

This is a preprint of INTERNETWARE'14.